\documentclass[a4paper,20pt]{article}
    \usepackage[T1]{fontenc}
    \usepackage{graphicx}
    \usepackage{epsfig}
    \usepackage{amsmath}
    \usepackage{amsfonts}
    \renewcommand{\abstract}{}
\begin{document}
\makeatletter
\renewcommand{\@evenfoot}{\hfil \thepage \hfil}
\renewcommand{\@oddfoot}{\hfil \thepage \hfil}
\fontsize{12}{12} \selectfont

\title{XMM-Newton Observations of X-ray Pulsar Cen X-3}
\author{\textsl{A.~V.~Tugay, A.~A.~Vasylenko}}
\date{}
\maketitle
\begin{center} {\small National Taras Shevchenko University of Kyiv, Glushkova ave., 2, 03127, Kyiv, Ukraine\\
                       tugay@univ.kiev.ua}
\end{center}

\begin{abstract}
We present the results of our study of X-ray pulsar Cen X-3 using XMM-Newton observations. The light curve and the spectrum for this observations were built and Fe triplet within $6.5-7$~keV region was detected.
The geometric model of relativistic accretion disk \cite{fabian89} for iron emission lines Fe~I 
$K_{\alpha}$,  Fe~XXV and Fe~XXVI in $6.4-7.0$~keV region was applied. The values of disc inclination, inner and outer radii of the disc and mass of the central compact object (neutron star) were obtained. Intensity variations of these lines during orbital motion were also detected. The largest variation was detected for Fe~I $K_{\alpha}$ line, that agrees with the results of other authors \cite{ebisawa96}. These results conform the model in which Fe~I $K_{\alpha}$ line forms in hot plasma of accretion disc and highly ionized iron lines form in outer regions of binary system. Probably the most interesting feature of Cen X-3 spectrum is Fe~XXV triplet which was found by Iaria~et al. \cite{iaria05} from Chandra data analysis. We did not find this triplet in our  analysis of XMM-Newton data and explain its absence by the insufficient energy resolution of XMM-Newton instruments. 
\end{abstract}

\section*{Introduction}
\indent \indent X-ray pulsars are binary systems consisting of bright massive star and neutron star. The last one draws a matter from a massive bright star and forms an accretion disk around itself. The accretion disk emits X-ray radiation within the range of energies from hundreds eV to tens keV. X-ray pulsar Cen X-3 is a pair of blue giant and neutron star at a distance of 8~kpc from the Earth. The mass of the neutron star is 1.2~$M_{\odot}$ and mass of the giant is 20~$M_{\odot}$. This object was discovered in 1967 with the help of sub-orbital rocket. X-ray pulses with period of 4.84~sec caused by neutron star spin were detected by UHURU satellite \cite{giacconi71} in 1971. The binary system was optically identified in \cite{krzeminski74}. The binary has eclipses with orbital period of 2.09~days. Out of eclipse phase the iron emission line in the spectrum of Cen X-3 is detected at 6.5~keV \cite{nagase92}. This source was observed by series of different modern high-energy telescopes: gamma-satellite INTEGRAL within the range of $16-60$~keV \cite{filippova05}, ground-based Cherenkov telescope Mark-6 at energies above 400~GeV \cite{atoyan02} and X-ray space observatories Chandra and XMM-Newton within the range of $0.2-12$~keV. Our goal was to process and analyse XMM-Newton data of Cen X-3. 

\section*{The data}
\indent \indent In our investigations we used the data obtained by XMM-Newton satellite on January 27, 2001 during 67~kiloseconds observation. These data are public available on HEASARC (High Energy Astrophysics Scientific Archive Resources Center) database web-page (http://heasarc.gsfc.nasa.gov). XMM-Newton satellite contains three EPIC cameras (MOS1, MOS2 and PN) that allows us to build the images, spectra and light curves for X-ray sources. During this observation two cameras, PN and MOS2, worked in imaging mode, and MOS1 camera worked in the light curve building mode (Timing mode). During observations in Timing mode time resolution and sensitivity of the camera increase, which is very useful for obtaining spectra and light curves. In this mode it is possible to build only histogram instead of two-dimensional image. For data processing SAS (Science Analysis System) shell ver.~1.52.8 was used.

During this observation, namely in its end, a solar flare appeared, which led to sharp rise of the light curve during 900 seconds. This time interval was excluded, and the resulting data contained 66~kiloseconds observation. For this last time interval we obtained an image using the data from PN camera, which has the greatest sensitivity and time resolution compared to other EPIC cameras. Another reason for choosing of PN camera is significant pile-up effect in MOS2 camera. This effect lies in the following. If two separate photons arrive during the time which is less than time resolution, then only one photon is detected by CCD matrix pixel instead of two and its energy is equal to the sum of energies of two photons. The influence of pile-up effect on spectrum is the following: firstly, continuum level increases at high energies and decreases at low energies; secondly, a part of photons is lost at all. 

\section*{Analysis of Light Curve and Spectrum}
\indent \indent In paper \cite{nagase92} the formula for calculation of Cen X-3 eclipses moments was derived taking into account the changes of system orbital period. Using this formula we determined the orbital phase of Cen X-3 system for the beginning and the end of XMM-Newton observation we consider. These phase values are -0.03 and 0.33. Phase 0 corresponds to the mid-eclipse and phases $\pm0.5$ correspond to the case when pulsar is in front of the star. The light curve of the observation for PN camera is presented in Fig.~1. One can see that X-ray flux significantly increases during post-eclipse phase.

To build the spectrum we select a circle region from the image with radius equal to 48~arcsec, thus the circle touches the image border. From the spectrum of this region the background spectrum was substracted. The background spectrum was taken from the rest part of the image. It is worth to note that we select the maximum possible resolution of the instrument, 15~eV, as the energy resolution of the spectrum. The resulting spectrum was handled by XSPEC software package ver.~12.2.

We excluded $0-0.2$~keV range, because XMM-Newton is not sensitive in this energy range, and obtained the spectrum, which is presented in~Fig.~2. This plot was built without taking into account any physical models of radiation. We fit the continuum level by the powerlaw model taking into account the photoelectric absorption along the line of sight \cite{balucinska92} and the iron absorption limit at 7.2~keV. Following \cite{iaria05} we used frozen values for spectral power-law index $\alpha =1.2$ and hydrogen column density $N_H =1.95 \cdot 10^{22}$~cm$^{-2}$. We obtained the values of normalization factor $N_{po} =(2.91 \pm 0.06) \cdot 10^{-2}$cm$^{-2}$~s$^{-1}$keV$^{-1}$, absorption limit energy $E_f =7.14\pm 0.03$~keV and optical depth $\tau =0.22\pm 3 $. In $6-7$~keV range of the spectrum one can see 3 iron emission lines. The first line of triplet is $K_{\alpha}$ line, which is emitted by neutral and low-ionised iron atoms. The second and the third lines are formed by helium-like ion Fe XXV and hydrogen-like Fe XXVI correspondingly. For more detailed study of these lines we rebuilt the spectrum for $6-7.5$~keV region. We fitted emission lines by Gaussian and Lorentzian profiles and found that the first one is more suitable. We obtained the central energies, widths and intensities for all lines and uncertainties of these parameters. The iron emission triplet can be seen in Fig.~3 and parameters of lines are presented in Table~1. In addition to Gaussian profile we used ``diskline'' model describing relativistic accretion disk \cite{fabian89}. This model allows us to estimate the angle between accretion disk and line of sight and the inner and outer radii of the accretion disk. We found that the sum of diskline and Gaussian profiles is needed to account the finite width of $K_\alpha$ line due to thermal broadening. We obtained the following inclination angles of the accretion disk: $60^{\circ}$ for K$\alpha$~line and $55^{\circ}$ for Fe~XXV and Fe~XXVI lines. The difference of inclination angle for $K_!
 {\alpha}
$~line and other lines is due to the fact, that it is formed in the outer atmosphere of the binary system. According to our model the accretion disk has inner radius of 4500~km and outer radius (by Fe~XXV and Fe~XXVI lines) of 405000~km. Our results agree within the errors with previous observations \cite{nagase89} according to which the inclination angle is equal to $75^{\circ} \pm 12^{\circ}$. The iron triplet with diskline approximation is presented in Fig.~4.

\section*{Comparison with other observations and conclusions}
\indent \indent Beside XMM-Newton Cen X-3 was observed by other satellites in particular Chandra \cite{iaria05} and ASCA \cite{ebisawa96}. These two satellites obtained Cen X-3 spectra in which iron lines in $6-7$~keV region were detected. We compared iron lines parameters from these observations with our results.

Chandra observed Cen X-3 on December 30, 2000 during of 48 kilosecoonds at orbital phases of $0.13-0.40$. Chandra group determined the equivalet width of $K_{\alpha}$ line and compared it with the results of \cite{ebisawa96} and \cite{wojdowski03}. We compared our $K_{\alpha}$ line parameters with Chandra data during post-eclipse phase which partially corresponds to XMM-Newton time of observations. The central energies of the line are the same for both satellites. Line intensity for XMM-Newton data is less than for Chandra: $(0.89\pm0.06) \cdot 10^{-3}$cm$^{-2}$s$^{-1}$ and $(1.18\pm0.19) \cdot 10^{-3}$cm$^{-2}$s$^{-1}$ correspondingly. This difference lies within the errors and is negligible. According to \cite{iaria05} $K_{\alpha}$ line width is equal to $11.5\pm4.5$~eV, which is 5 times less than for XMM-Newton observation. We fixed $K_{\alpha}$ line width at 11~eV and obtained increasing of the $\chi^2/d.o.f.$ value from 1.0126 to 1.2463 and corresponding F-statistics $F=66$. This means that models with 11~keV and 52~keV are similar with probability of $10^{-14}$. Thus we concluded that for XMM-Newton observation intrinsic line width is significantly larger and this effect can not be explained by the differences in energy resolution of XMM-Newton and Chandra detectors. Chandra observed Cen X-3 when it was in outburst state while the time of XMM-Newton observation (4 week later) corresponds to quiescent state of system. Thus the differences in X-ray emission can be explained by the different states of Cen X-3. The values of continuum level and power-law normalization factor for XMM-Newton data is 19 times less that also agrees with quiescent state.

It is clearly seen from the light curve that the source brightness varies significantly with the orbital phase. We considered the changes of the intensity of iron lines with phase. This effect was discussed at first time in \cite{ebisawa96}. The authors examined the changes of Cen X-3 iron lines using ASCA X-ray satellite. This satellite observed Cen X-3 during June $24-25$, 1993 at the following orbital phases: $-0.31/-0.27$ (pre-eclipse), $-0.23/-0.08$ (ingress), $-0.07 / 0.12$ (eclipse), $0.14 / 0.19$ (egress). XMM-Newton observation interval corresponds to a part of ASCA eclipse interval and the whole egress interval. For these two phase intervals we built the source spectra from XMM-Newton observation and found the parameters of iron lines (see Table~1). We compared parameters of iron lines for XMM-Newton and ASCA in eclipse and egress phases. Taking into account the difference of effective areas of satellites detectors we found that continuum levels within 6.5~keV region are equal. Intensities of all lines from XMM-Newton observation coincides with ASCA data within errors. We confirmed the orbital variations of iron lines. Positions of the lines also coincide within errors. When comparing the equivalent widths of the lines we revealed the following. During eclipse phase Fe~XXVI line is two times broader for XMM-Newton than for ASCA, but two other lines coincide within errors. During egress phase all lines from XMM-Newton observation have 1.5~times larger width than from ASCA. These differences depend on the features of Cen X-3 accretion state at the time of ASCA observation (June $24-25$, 1993). 

Therefore the main comparison results are the confirmation of orbital variations, identified by the group of Japanese satellite ASCA \cite{ebisawa96} and quantitative conformity of lines parameters, namely positions and intensities.

In \cite{iaria05} the discovery of Fe~XXV triplet instead of single line was reported. The components of triplet were interpreted as forbidden transition, intercombination and resonance lines. In XMM-Newton spectrum we can see only single Fe~XXV line with the parameters given in Table~1. This gives us an evidence that Fe~XXV triplet could be resolved only at high state of binary system. Also we found Fe~XXVI line at 6.9~keV, which was observed by ASCA group but is absent in object spectrum from Chandra. We found that lines parameters, namely their orbital variations, position and intensity, obtained from ASCA and XMM-Newton observations coincide. In \cite{iaria05} the authors explained the lack of Fe~XXVI line by the stellar wind decreasing. Taking into account high state of Cen X-3 during Chandra observation we conclude that stellar wind rate is not the crucial factor that affects X-ray luminosity of the system. 

\begin{table*}
\caption{Parameters of Cen X-3 spectrum. Parameters in columns 2 and $3-5$ were obtained for the whole energy interval ($0.2-12$~keV) and $5-8$~keV interval correspondingly. Normalization parameter is measured in units of $10^{-5}$~cm$^{-2}$s$^{-1}$keV$^{-1}$ for iron emission lines and in $10^{-3}$~cm$^{-2}$s$^{-1}$keV$^{-1}$ for power-law model for the whole spectrum.}
\centering
\begin{tabular}{|c|c|c|c|c|c|}
\hline
Parameter & Full XMM observation & Chandra egress & ASCA eclipse & ASCA egress \\
\hline
Norm & $29.1\pm5.6$ & $43.0\pm0.4$ & $6.5\pm0.3$ & $34.0\pm7.6$ \\
\hline
Line energy ($K_{\alpha}$) & $6386\pm5$ & $6394\pm3$ & $6377\pm15$ & $6397\pm7$ \\
\hline
Sigma & $36\pm15$ & $52\pm5$ & $76\pm21$ & $49\pm15$ \\
\hline
Norm & $49\pm7$ & $89\pm6$ & $11\pm4$ & $60\pm10$ \\
\hline
Line energy (Fe XXV) & $6654\pm5$ & $6668\pm3$ & $6655\pm5$ & $6666\pm6$ \\
\hline
Sigma & $46\pm14$ & $46\pm8$ & $45\pm14$ & $44\pm17$ \\
\hline
Norm & $53\pm7$ & $78\pm6$ & $24\pm3$ & $65\pm10$ \\
\hline
Line energy (Fe XXVI) & $6945\pm7$ & $6969\pm3$ & $6941\pm5$ & $6966\pm7$ \\
\hline
Sigma & $43\pm19$ & $53\pm8$ & $37\pm13$ & $56\pm15$ \\
\hline
Norm & $37\pm7$ & $61\pm5$ & $21\pm3$ & $52\pm9$ \\
\hline
Reduced $\chi ^2 $ & 1.0184 & 1.0126 & 0.9115 & 1.0851 \\
\hline
\end{tabular}
\end{table*}

\section*{Acknowledgement}
This work was done in the Ukrainian Virtual Roentgen and Gamma Observatory VIRGO.UA.

\begin{figure}[p]
\centering
\epsfig{figure=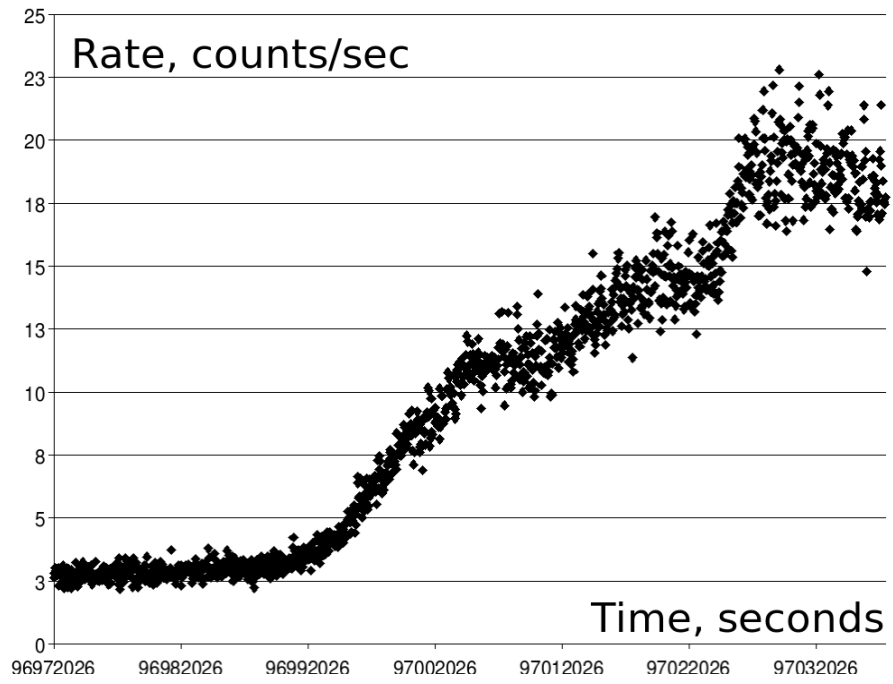,width=.65\linewidth}
\caption{Light curve of Cen X-3 for XMM-Newton observation. Time counts from January 1, 1998.}
\label{fig1}
\end{figure}

\begin{figure}[p]
\centering
\epsfig{figure=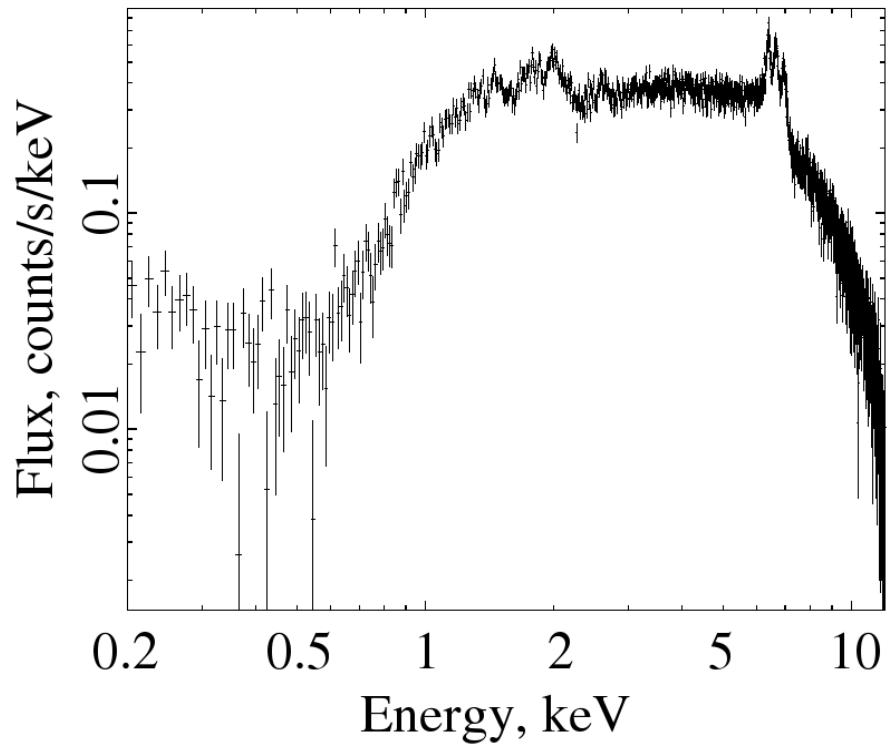,width=.65\linewidth}
\caption{Cen X-3 spectrum from XMM-Newton.}
\label{fig2}
\end{figure}

\begin{figure}[p]
\centering
\epsfig{figure=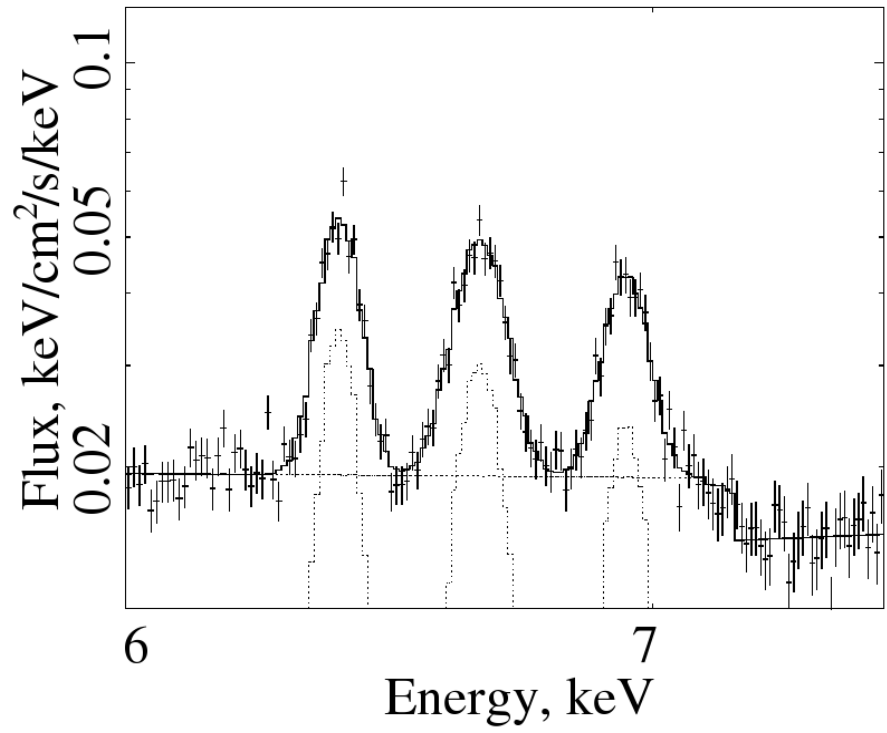,width=.65\linewidth}
\caption{Iron emission lines fitted by Gaussian model.}
\label{fig2}
\end{figure}

\begin{figure}[p]
\centering
\epsfig{figure=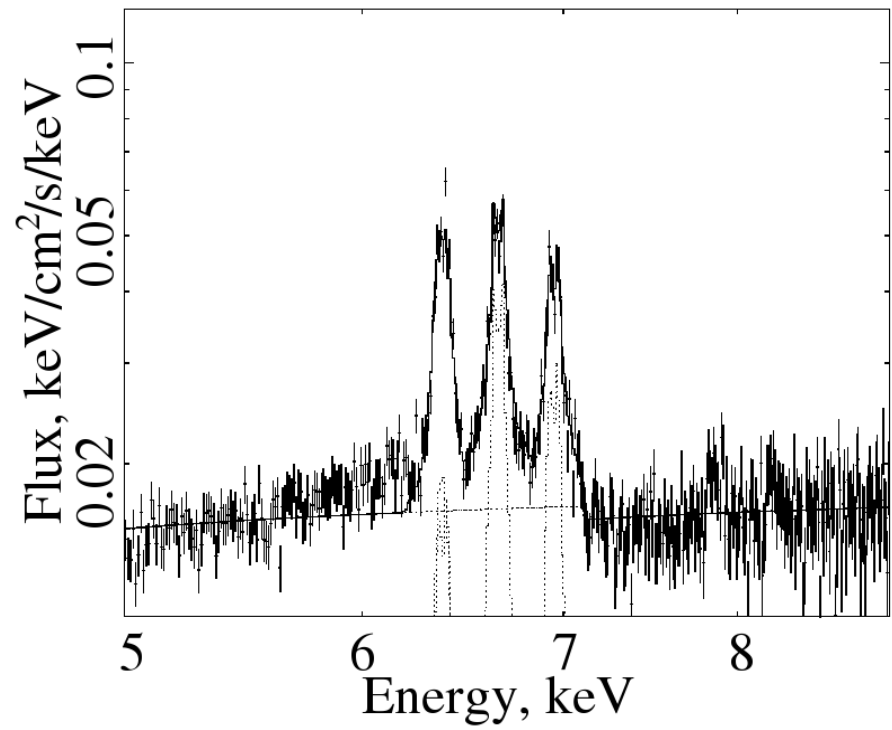,width=.65\linewidth}
\caption{Iron emission lines fitted by diskline model.}
\label{fig2}
\end{figure}

\end{document}